\documentclass[twocolumn]{aastex631}

\usepackage{CJK}
\usepackage[T1]{fontenc} 
\usepackage[utf8]{inputenc}

\usepackage{amsfonts}
\usepackage{amsmath}
\usepackage{mathrsfs}
\usepackage{epsfig}
\usepackage{subfigure}
\usepackage{multirow}
\usepackage{threeparttable}

\def\BJD{\,\mathrm{BJD}}

\def\Porb{\,P_\mathrm{orb}}

\def\cd{\mathrm{c/d}}
\def\Msun{\,{M}_{\odot}}

\def\raa{Res.\ Astron.\ Astrophys.\ }

\shorttitle{BE Lyn: A Star in the Most Eccentric Binary}
\shortauthors{Niu, Zhang \& Xue}

\begin{document}
\begin{CJK*}{UTF8}{gbsn}

\title{BE Lyncis: A Pulsating Star in the Most Eccentric Binary with a Massive Unseen Companion}

\author[0000-0001-5232-9500]{Jia-Shu Niu (牛家树)}
\affil{Institute of Theoretical Physics, Shanxi University, Taiyuan 030006, China}
\affil{State Key Laboratory of Quantum Optics Technologies and Devices, Shanxi University, Taiyuan 030006, China}
\affil{Collaborative Innovation Center of Extreme Optics, Shanxi University, Taiyuan 030006, China}
\email{jsniu@sxu.edu.cn}
\correspondingauthor{Jia-Shu Niu}

\author{Ying Zhang (张颖)}
\affil{Institute of Theoretical Physics, Shanxi University, Taiyuan 030006, China}
\email{}

\author[0000-0001-6027-4562]{Hui-Fang Xue (薛会芳)}
\affil{Department of Physics, Taiyuan Normal University, Jinzhong 030619, China}
\affil{Institute of Computational and Applied Physics, Taiyuan Normal University, Jinzhong 030619, China}
\affil{Shanxi Key Laboratory for Intelligent Optimization Computing and Blockchain Technology, Jinzhong 030619, China}
\email{hfxue@tynu.edu.cn}
\correspondingauthor{Hui-Fang Xue}

\begin{abstract}
We report the discovery of an exceptionally eccentric binary system, BE Lyncis (BE~Lyn), which might host a compact companion with mass $\gtrsim 2.5~M_{\odot}$. By combining \textit{TESS} photometry with an extensive set of times of maximum light spanning 39~years, we identify BE~Lyn as a high-amplitude $\delta$ Scuti star in a binary with an orbital period of $\approx15.9$~years and an extraordinary eccentricity of $e=0.9989^{+0.0008}_{-0.0021}$ ($>0.9968$ at 95\% confidence)---the most extreme eccentricity reliably measured for any binary system. Dynamical constraints limit the orbital inclination to $i \lesssim 10.1^{\circ}$, implying a companion mass $M_2 \gtrsim 2.5~M_{\odot}$, which identifies the companion as a compact object. This mass points to it most likely being a black hole; if instead it is a rapidly rotating neutron star, it would be the most massive known. If the black hole interpretation holds, it would be the closest such object to Earth. This system provides a unique laboratory for studying asteroseismology in strong gravitational fields, as well as the formation and evolution of extremely eccentric binaries. Our work demonstrates the use of the light-travel time effect in a pulsating star to reveal a compact companion, offering a novel method for detecting black holes in non-interacting binaries.
\end{abstract}

\section{Introduction}
\label{sec:intro}

Pulsating $\delta$ Scuti stars, with spectral types A--F, occupy the intersection of the main sequence and the classical Cepheid instability strip in the Hertzsprung--Russell diagram. Their oscillations, driven by the $\kappa$ mechanism operating in the helium partial ionization zone, provide stable periodic signals with periods ranging from 15 minutes to 8 hours \citep{Kallinger2008, Handler2009, Guenther2009, Uytterhoeven2011, Holdsworth2014, Steindl2022}. High-amplitude $\delta$ Scuti stars (HADS), a subclass exhibiting photometric variations exceeding 0.1 mag and typically slow rotation ($v\sin i \leq 30$ km s$^{-1}$), serve as precision celestial clocks. Their stability makes them ideally suited for detecting the minute period modulations induced by orbital motion in binary systems---a phenomenon known as the light-travel time effect (LTTE).

BE Lyncis (BE~Lyn; HD~79889) is a well-studied HADS with a mean magnitude of $\langle V\rangle = 8.8$ mag, spectral type A3, and sub-solar metallicity ([Fe/H] = $-0.598$) \citep{Kim_2012, Abdurrouf2022}. First identified as a variable star by \citet{Oja_1986} and later confirmed as a HADS with a pulsation period of $\sim0.09587$ days \citep{Oja_1987}, it has been the subject of numerous period variation studies. Anomalies in its $O-C$ diagram have long hinted at the presence of a binary companion, occasionally manifesting as a parabolic trend \citep[e.g.,][]{Rodriguez1990, Liu_1991, Wunder_1992, Boonyarak2011}. Several investigations indeed proposed the existence of a low-mass companion \citep{Kiss_1995, Derekas2009, Boonyarak2011, Pena2015, Liakos2017}, though the inferred orbital parameters varied substantially owing to limited temporal coverage and data precision.

Two key developments now enable a definitive reassessment of this system. First, high-precision, continuous photometry from the Transiting Exoplanet Survey Satellite (\textit{TESS}) allows exquisite characterization of the primary star's pulsation modes and yields significantly more accurate times of maximum light. Second, modern computational techniques, such as robust multi-parameter global fitting via Markov Chain Monte Carlo (MCMC) methods, permit a physically consistent and precise determination of binary orbital parameters.

Here we present a comprehensive analysis of BE~Lyn, combining \textit{TESS} photometry with a newly compiled, 39-year baseline of times of maximum light. Our work not only confirms its binary nature but reveals an astrophysical gem: an ultra-wide, extremely eccentric orbit harbouring a compact companion with a mass $\gtrsim 2.5~M_{\odot}$,   making it a strong black hole candidate. If confirmed as such, it would be the closest black hole to Earth, at a distance of approximately 250 pc \citep{Bailer2021}. 

The system holds multiple distinctions: (i) it possesses the highest orbital eccentricity ever reliably measured in any binary; (ii) it is the only known binary comprising a pulsating star and a massive compact companion, opening a new window for asteroseismology in strong gravitational fields; (iii) its extreme eccentricity ($e = 0.9989$) provides a stringent observational test for models of supernova natal kicks and post-supernova mass transfer, offering critical insights into the formation pathways of such extreme binaries; and (iv) it validates the LTTE method using pulsating stars as a novel, independent pathway for discovering quiescent compact objects in non-interacting binaries. BE~Lyn thus stands as a benchmark system, with profound implications for stellar evolution, compact-object demographics, and strong-gravity astrophysics.

\section{Results and Discussion}
\label{sec:results_discussion}

\subsection{Pulsation Properties and the Stellar Clock}
\label{subsec:pulsation}

Analysis of the \textit{TESS} photometry reveals BE~Lyn as a multi-periodic pulsator. We identify 32 significant frequencies, comprising 7 independent modes and 25 harmonics or combination frequencies (see Appendix~\ref{app:photometry} for more details). The two dominant frequencies, $f_0 = 10.4309 \pm 0.0005$ c/d and $f_1 = 13.4226 \pm 0.0005$ c/d, exhibit a period ratio $P_1/P_0 = 0.777$, unambiguously identifying them as the fundamental and first-overtone radial modes, respectively \citep{Xue2023}. The amplitude of the fundamental mode dominates the light curve, exceeding that of the first overtone by a factor of $\sim45$ and those of the remaining independent modes by factors $>700$. This clear amplitude hierarchy makes the $f_0$ pulsation an exceptionally clean and stable clock, enabling the classical $O-C$ method to detect minute orbital perturbations with negligible contamination from multi-periodic effects. The five additional independent frequencies ($f_2$--$f_6$), which lie in the $9-17$ c/d range with low amplitudes, are likely non-radial p-modes. This rich pulsation spectrum establishes BE~Lyn as a promising target for future detailed asteroseismic modeling.

\subsection{Orbital Solution: An Extraordinarily Eccentric Binary}
\label{subsec:orbit}

\begin{figure*}[!htbp]
    \centering
    \includegraphics[width=0.8\textwidth]{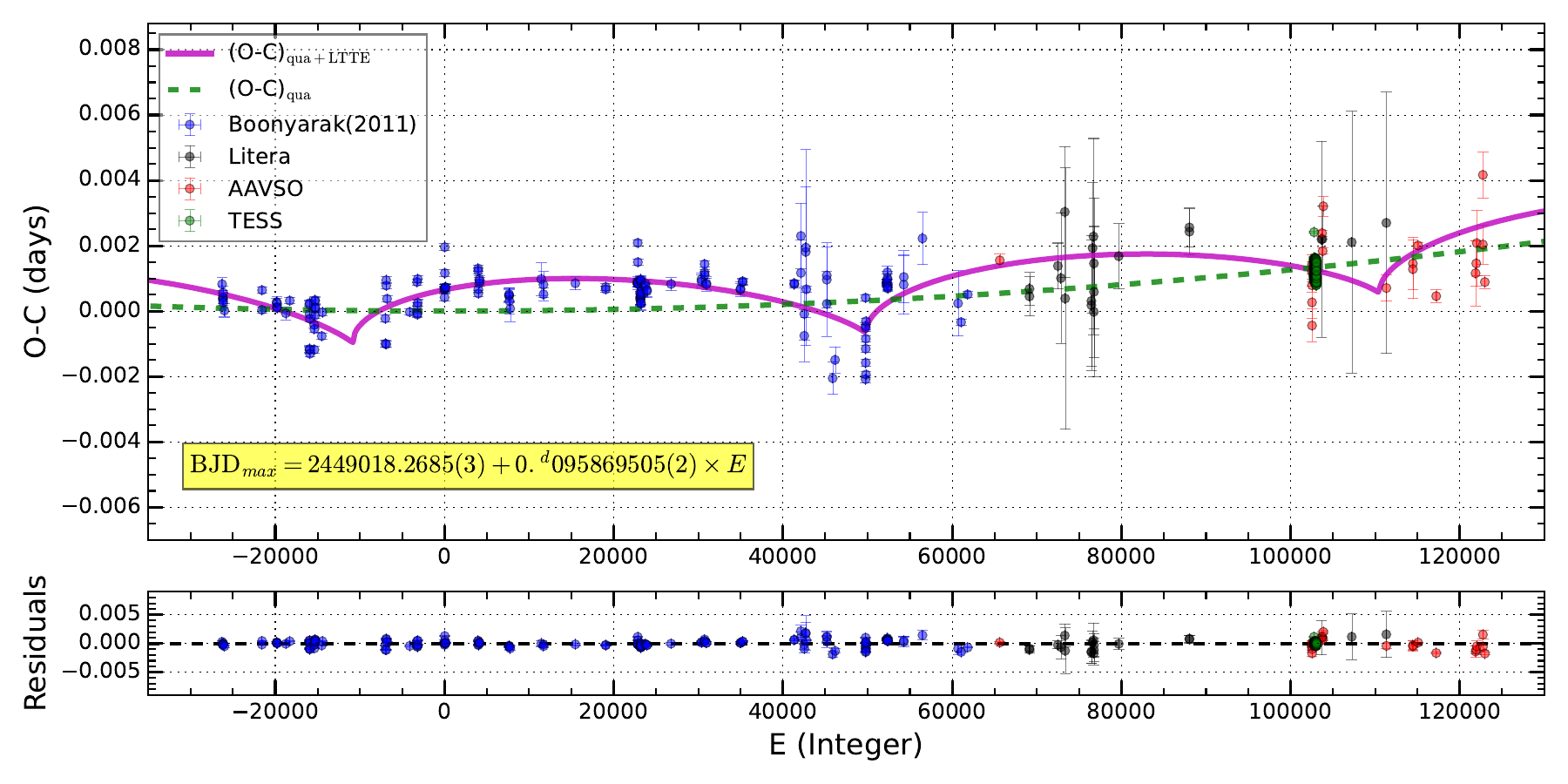}
    \caption{$O-C$ diagram of BE Lyn (with the linear ephemeris subtracted). The top panel shows the data and the best-fit model (magenta solid line), with the quadratic component indicated by the green dashed line. The bottom panel displays the residuals. Data sources are color-coded: \citet{Boonyarak2011} (blue), historical literature \citep{Hubscher_2013, Hubscher_2014, Hubscher_2017,Pena2015, Pagel_2021, Pagel_2022, Pagel_2023} (gray), AAVSO (red), and TESS (green).}
    \label{fig:OC_BELyn}
\end{figure*}

The $O-C$ analysis, incorporating a 39-year baseline of 442 times of maximum light, reveals a pronounced long-period modulation superimposed on a secular period increase (Figure~\ref{fig:OC_BELyn}). Global fitting with a model combining a quadratic term (intrinsic period variation) and a light-travel time effect (LTTE) term yields the orbital solution presented in Table~\ref{tab:pul_orb_para}.

\begin{deluxetable}{cc}[!htbp]
    \label{tab:pul_orb_para}
    \tablecaption{Pulsation and Orbital Parameters for BE Lyn.}
    \tabletypesize{\footnotesize}
    \tablehead{
        \colhead{Parameter} & \colhead{Value}
    }
    \startdata
    $\BJD_0$ & $2449018.26853^{+0.00002}_{-0.00002}$ \\
    $P_0$ (d) & $0.0958695052^{+0.0000000009}_{-0.0000000009}$ \\
    $\beta$ (d~cycle$^{-1}$) & $(2.6^{+0.2}_{-0.2})\times10^{-13}$ \\
    $A$ (d) & $0.006^{+0.007}_{-0.003}$ \\
    $e$ & $0.9989^{+0.0008}_{-0.0021}$ \\
    $\Porb$ (d) & $5811^{+29}_{-30}$ \\
    $T_0$ (BJD) & $2447989^{+30}_{-30}$ \\
    $\omega$ ($^{\circ}$) & $351^{+5}_{-6}$ \\ \hline
    $(1/P_0)(dP_0/dt)$ (yr$^{-1}$)\tablenotemark{a} & $(1.02^{+0.08}_{-0.08})\times10^{-8}$ \\
    $a \sin i$ (AU)\tablenotemark{a} & $1.1^{+1.3}_{-0.4}$ \\
    $f(m)$ ($\Msun$)\tablenotemark{a} & $0.005^{+0.046}_{-0.004}$ \\
    \enddata
    \tablecomments{
        \tablenotetext{a}{Derived from the directly fitted parameters above.}
    }
\end{deluxetable}

The most striking result is the extraordinary orbital eccentricity: $e = 0.9989^{+0.0008}_{-0.0021}$ ($>0.9968$ at 95\% confidence). With this value, BE~Lyn sets a new record for the highest reliably measured eccentricity in any binary system, surpassing previous benchmarks such as WDS 05354-3316 ($e=0.985\pm0.002$ \citep{Tokovinin2020}) and $\zeta$ Boo ($e=0.980450\pm0.000064$ \citep{Waisberg_2024, Waisberg_2025}). The orbital period is $P_{\mathrm{orb}} = 5811^{+29}_{-30}$ days ($\approx 15.9$ years). The associated mass function, $f(m) = 0.005^{+0.046}_{-0.004} M_{\odot}$, and small projected semi-major axis ($a\sin i = 1.1^{+1.3}_{-0.4}$ AU) point to a massive, unseen companion. This discovery highlights the power of long-term, precise photometric monitoring of pulsating stars to uncover extreme binary architectures.

\subsection{Asteroseismic Properties of the Primary Star}
\label{subsec:primary}

\begin{figure}[!htbp]
    \centering
    \includegraphics[width=0.48\textwidth]{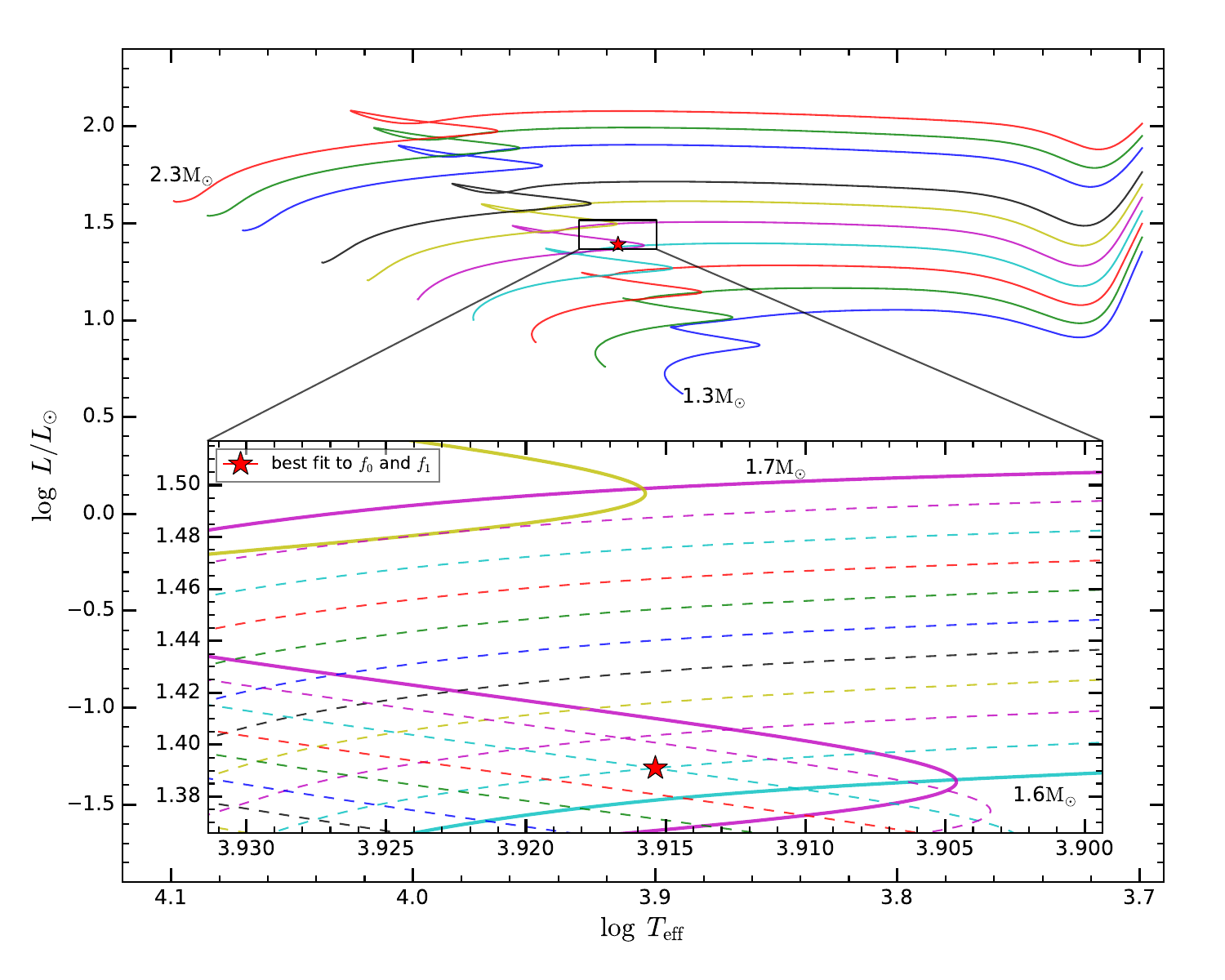}
    \caption{Hertzsprung--Russell diagram showing evolutionary tracks (colored solid lines from the zero-age main sequence, ZAMS, to post-MS) and the best-fit seismic model for BE~Lyn's primary (red star). Black rectangles highlight zoomed-in views around the best-fit solutions. Colored dashed lines indicate tracks with a $0.01 M_{\odot}$ mass step.}
    \label{fig:hr}
\end{figure}

To place dynamical constraints in context, we determined the properties of the primary star through asteroseismic modeling. We computed a grid of stellar evolutionary tracks using MESA \citep[Modules for Experiments in Stellar Astrophysics;][]{Paxton2011, Paxton2013, Paxton2015, Paxton2018, Paxton2019, Jermyn2023} and corresponding adiabatic pulsation frequencies using GYRE \citep{Townsend2013, Townsend2018, Goldstein2020, Sun2023}. The optimal model, selected by matching the observed frequencies $f_0$ and $f_1$, is summarized in Table~\ref{tab:primary_para} and located on the Hertzsprung--Russell diagram in Figure~\ref{fig:hr}.

The best-fit model places the primary on the post-main sequence, with a mass of $M_1 = 1.61 M_{\odot}$, effective temperature $\log T_{\mathrm{eff}} = 3.915$, and luminosity $\log (L/L_{\odot}) = 1.391$. Notably, the observed period variation rate, $(1.02 \pm 0.08) \times 10^{-8}$ yr$^{-1}$, differs from the model-predicted rate of $8.36 \times 10^{-8}$ yr$^{-1}$. This discrepancy may indicate ongoing mass transfer during the observed time interval.

The radius derived from the best-fit asteroseismic model (approximately $2.4\,R_{\odot}$, see Table~\ref{tab:primary_para}) is considered to be an important parameter of the primary star, which will provide a significant constraint on the binary system (see in next subsection).

\subsection{A Massive Compact Companion in an Extreme Orbit}
\label{subsec:companion}

\begin{figure}[!htbp]
    \centering
    \includegraphics[width=0.48\textwidth]{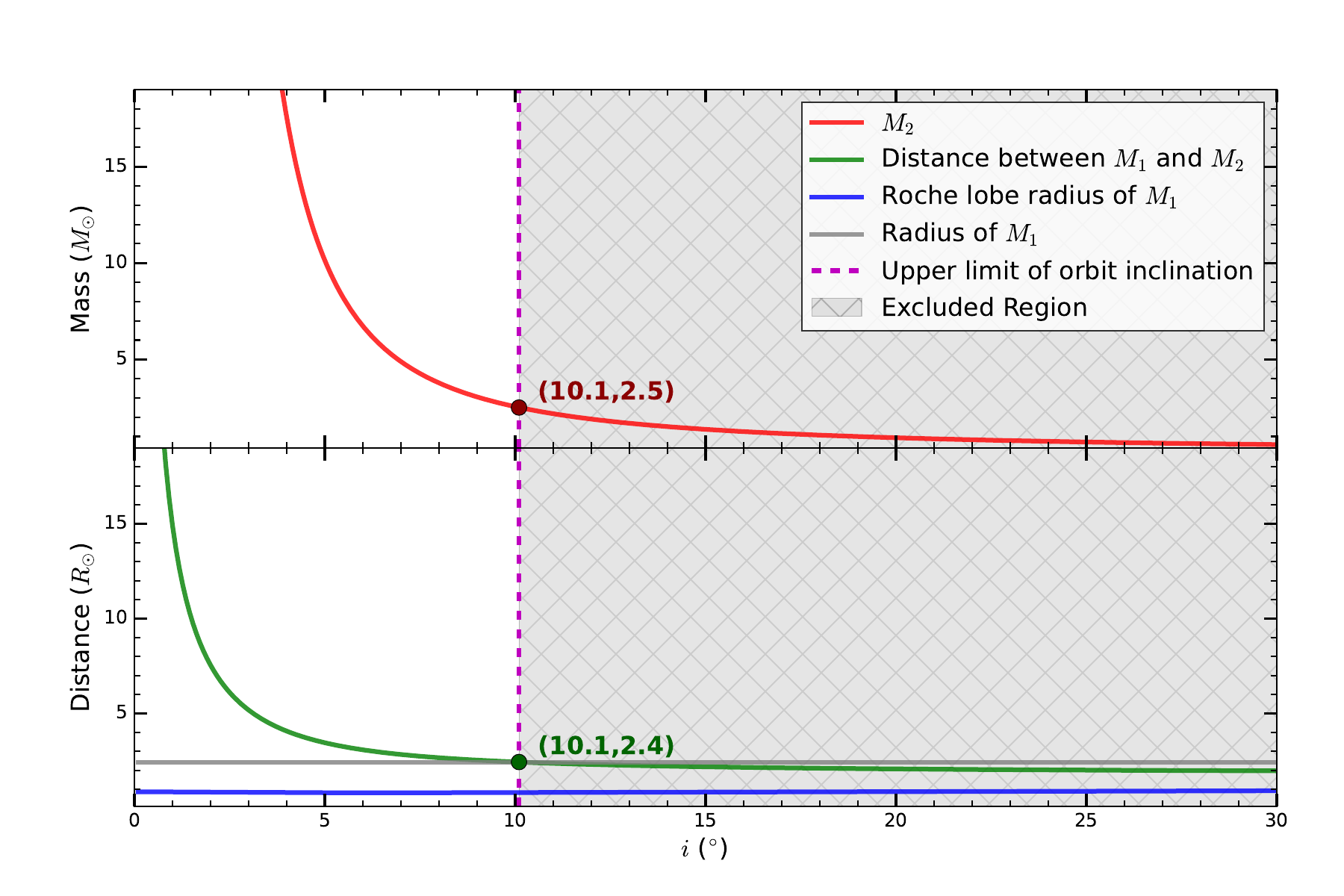}
    \caption{Dynamical constraints on the unseen companion. \textit{Top:} Companion mass $M_2$ versus orbital inclination $i$, derived from the mass function (red curve). \textit{Bottom:} Primary-companion separation at periastron versus $i$ (green curve). The intersection of the primary's radius $R_{1}$ (grey curve) and primary-companion separation (green curve) sets the stringent upper limit $i \lesssim 10.1^\circ$ (dark green dot), corresponding to the lower mass limit $M_2 \gtrsim 2.5 M_{\odot}$ (dark red dot). The gray hatched region denotes parameter space excluded.}
    \label{fig:m2_mass}
\end{figure}

Combining the orbital parameters with the primary mass ($M_1 = 1.61 M_{\odot}$) introduces a critical constraint from orbital stability: the periastron separation must exceed the primary's radius to avoid catastrophic collision. This condition tightly constrains the system geometry (Figure~\ref{fig:m2_mass}). We derive an upper limit on the orbital inclination of $i \lesssim 10.1^\circ$, implying a lower limit on the companion mass of $M_2 \gtrsim 2.5 M_{\odot}$.

While a periastron separation smaller than the tidal disruption radius $R_1 (M_2/M_1)^{1/3}$ might appear to threaten the system's integrity, a more careful assessment is required. The relevant timescale for the star to respond to tidal forces is its dynamical timescale $t_{\rm dyn} = \sqrt{R_1^3/(G M_1)} \approx 1.3$ hr. This is comparable to the $\approx 0.92$ hr interval during which the orbital separation falls below the tidal disruption radius for the limiting case $i = 10.1^\circ$ (see Figure~\ref{fig:m2_mass}). Thus, the star may not be able to maintain hydrostatic equilibrium during periastron passage, and partial or even complete tidal disruption is physically plausible.

A more quantitative criterion is given by the penetration factor $\beta \equiv R_{\rm tide} / r_{\rm peri}$, where $R_{\rm tide} = R_1 (M_2/M_1)^{1/3}$ and $r_{\rm peri}$ is the periastron separation. For $i = 10.1^\circ$, we find $\beta \approx 1.16$. According to the simulations of \citet{Guillochon2013}, for a star with a convective envelope (polytropic index $\gamma = 5/3$), $\beta \gtrsim 1.1$ leads to complete disruption; for more centrally concentrated stars (lower $\gamma$), the core may survive but a significant fraction of the envelope can be stripped during each passage.
Considering that BE Lyn is a post-MS star, whose 90\% mass is concentrated within 30\% of its radius (based on its best-fit asteroseismic model), the system could experience partial tidal disruption events repeatedly every $\approx 16$ yr.

Such repeated micro-tidal disruption events would be expected to leave observable imprints. Over the 39 yr observational baseline, they might cause quasi-periodic brightness variations, changes in the pulsation properties, or the build-up of a detectable accretion signature. Due to the lack of high-precision continuous photometric data covering the vicinity of periastron, the available TESS light curve and the extensive collection of times of maximum light do not show obvious outbursts or abrupt changes in the pulsation period, but deeper multi-wavelength observations (UV/X-ray) are needed to constrain such processes. Future monitoring will be essential to determine whether BE Lyn is indeed a repeating partial tidal disruption candidate.

While natal kicks during an asymmetric supernova explosion can impart a relatively high eccentricity to binary systems \citep{Janka2013, Popov2025}, the extreme eccentricity and semi-major axis of this system may require additional mechanisms to explain its later evolution. Interestingly, at periastron, the primary's radius inevitably exceeds its Roche lobe radius (using the approximation of \citet{Eggleton1983}; see Figure~\ref{fig:m2_mass}), indicating that the primary must fill its Roche lobe, leading to mass transfer near periastron.\footnote{Here, we thank \citet{Nagarajan2026} for pointing out an error in our previous calculation of the primary's Roche lobe radius.} This periastron mass transfer may have driven the elongation of the semi-major axis and the extreme eccentricity \citep{Parkosidis2026a,Parkosidis2026b}, consistent with their unified semi-analytical framework and representing an extreme case of such evolution.

A compact object with mass $M_2 \gtrsim 2.5 M_{\odot}$ cannot be supported by electron degeneracy pressure, ruling out a white dwarf. 
Considering the maximal gravitational mass $M_\mathrm{TOV} = 2.25^{+0.08}_{-0.07}\ M_{\odot}$ of nonrotating neutron stars from \citet{Fan2024}, the companion could most likely be a black hole.

If the companion is a neutron star (rapidly rotating), it would challenge the current record for the most massive neutron star (PSR J0952-0607 at $2.35 \pm 0.17 M_{\odot}$ \citep{Romani2022}). If instead it is a black hole, it would become the nearest known black hole to Earth, at a \textit{Gaia} DR3 distance of approximately 250 pc \citep{Bailer2021}---roughly half the distance of the previous record-holders Gaia BH1 (480 pc \citep{El-Badry2023}) and Gaia BH3 (590 pc \citep{Gaia-BH3-2024}). Its proximity offers unparalleled opportunities for detailed follow-up observations, including astrometric monitoring and multi-wavelength searches for accretion signatures.

\subsection{Proper Motion Anomaly from Hipparcos and Gaia}
\label{subsec:PMA}

Upon submitting the first version of this work \citep{Niu2026} to arXiv, we received prompt and valuable feedback from \citet{Nagarajan2026}, who pointed out that the observed proper motion anomaly (PMA) over the 25-year baseline from Hipparcos and Gaia ruled out the existence of a massive companion. We thank them for identifying an error in our previous calculation of the primary's Roche lobe radius; correcting this rendered mass transfer at periastron inevitable and led to a revised lower limit for the companion mass ($M_2 \gtrsim 2.5 M_{\odot}$). However, the PMA issue warrants a more detailed examination here.
 
\citet{Nagarajan2026} find that the predicted proper motion anomaly (PMA) over the 25 yr baseline between Hipparcos and Gaia is at least an order of magnitude larger than the observed value of $\approx 1.7 \pm 0.8\ \mathrm{mas\,yr^{-1}}$, derived from the two missions \citep{Hipparcos,Gaia2018,Gaia2023}. However, the parallax obtained by Hipparcos for BE Lyn is $0.45 \pm 1.00\ \mathrm{mas}$ \citep{Hipparcos}, which shows a significant discrepancy with the Gaia DR3 value of $3.97 \pm 0.02\ \mathrm{mas}$ \citep{Gaia2023}. While this indicates that the Hipparcos parallax is unreliable, its proper motion measurements are robust, being averaged over the four-year mission lifetime. Indeed, the Hipparcos proper motions agree with those from Gaia DR3 within 2$\sigma$, implying that no significant PMA is detected over the 25 yr baseline (see also \citet{Brandt2021}). This constitutes an important observational constraint on the binary model.

Furthermore, the primary star's motion reverses very rapidly around periastron, an event lasting only about 3 hours (corresponding to a true anomaly range from $270^\circ$ to $90^\circ$). The time of periastron passage, $T_0$ in Table \ref{tab:pul_orb_para}, has an uncertainty of approximately 30 days. While this uncertainty affects the precise prediction of the star's motion at any given moment, the Hipparcos proper motions average over four years, so the predicted average proper motion over the mission interval is not dominated by this uncertainty. Consequently, the absence of a detectable PMA remains a meaningful test of the binary hypothesis.

BE Lyn is not included in the Gaia DR3 non-single star catalog, and its Renormalised Unit Weight Error (RUWE) is $1.07$ \citep{Gaia2023}. It might be tempting to argue that the photocenter motion is too slow and nearly rectilinear to be detected by Gaia. However, forward modeling with the \texttt{gaiamock} tool \citep{El-Badry2024} shows that for the binary parameters derived from our $O-C$ analysis (a $\gtrsim 2.5\,M_{\odot}$ companion), the astrometric signature over Gaia’s $\approx 2.8$ yr baseline would be sufficiently large to trigger a 7-parameter acceleration solution and yield RUWE $> 1.4$ \citep{Nagarajan2026}. The fact that BE Lyn has RUWE $=1.07$ and was not flagged as a non-single star therefore provides independent evidence against the presence of a massive compact companion. As with the PMA, this constraint will be sharpened with future Gaia data releases.

\subsection{Broader Implications and Future Prospects}
\label{subsec:implications}

BE~Lyn is a system of singular importance across multiple astrophysical frontiers. First, it is the only known binary comprising a pulsating star and a quiescent compact object spanning the mass gap. This unique configuration opens the door to \textit{strong-gravity asteroseismology}: in principle, the primary's pulsations could probe the local spacetime metric perturbed by the massive companion, especially during the brief, close periastron passages.

Second, this discovery validates the pulsation-timing method as a powerful and independent technique for detecting dormant stellar-mass compact objects in wide binaries. Complementary to radial-velocity and astrometric surveys, this approach is particularly sensitive to long-period systems where other methods require extended baselines.

Several key questions remain. Precise dynamical mass measurements via future astrometric missions (e.g., \textit{Gaia}, \textit{Theia}) or very-long-baseline interferometry could determine $M_2$ and $i$ unambiguously. Multi-wavelength searches for accretion signatures or bow-shock emission are warranted, despite the wide separation. Theoretical work is urgently needed to model binary evolution pathways that produce such extreme eccentricities post-supernova.

In conclusion, BE~Lyn is not merely another binary system; it is a benchmark object. It challenges our understanding of compact-object formation kicks, offers a new tool for discovering compact objects, and provides a unique cosmic laboratory at our galactic doorstep.

Finally, we note that despite the compelling interpretation of the $O-C$ variations as a LTTE, the binary model faces significant observational challenges, particularly from the absence of a proper motion anomaly and the lack of a Gaia astrometric signal. While alternative explanations (e.g., stochastic period fluctuations \citep{Nagarajan2026} or an undetected third component \citep{Lian2025}) cannot be ruled out at present, future Gaia data releases and multi‑wavelength campaigns will be essential to definitively confirm or refute the existence of a compact companion.

\section{Conclusion}
\label{sec:conclusion}

We have established BE~Lyncis as an extraordinary binary system. The primary is a post-main-sequence, high-amplitude $\delta$ Scuti star with a mass of $1.61\,M_{\odot}$. Its unseen companion follows an orbit with an eccentricity of $e = 0.9989^{+0.0008}_{-0.0021}$ (95\% confidence $>0.9968$), the most extreme reliably measured in any binary system to date. Dynamical constraints, rooted in the binary system's stability, limit the inclination to $i \lesssim 10.1^\circ$ and demand a companion mass $M_2 \gtrsim 2.5\,M_{\odot}$. This firmly establishes the companion as a compact object and, given current understanding of neutron star maximum mass, points to it most likely being a black hole.

The discovery of BE~Lyn carries profound implications across several astrophysical frontiers. First, its record-breaking eccentricity provides a stringent observational benchmark for models of natal kicks in asymmetric supernovae, challenging our understanding of compact-object formation and evolution. Second, it represents the first known binary system to pair a pulsating star with a quiescent compact object spanning the mass gap, opening the door to strong-gravity asteroseismology—a novel probe of spacetime around massive companions. Third, our work validates the pulsation-timing method, which harnesses the precise clocks of $\delta$ Scuti stars, as an independent and powerful pathway to discover dormant compact objects in wide binaries. This technique complements existing radial-velocity and astrometric surveys and offers a new observational window into the hidden population of neutron stars and black holes.

Fundamental questions remain. The exact mass and nature of the companion—whether the most massive neutron star known or the nearest black hole to Earth—await definitive measurement. The evolutionary pathway that produced such an extreme configuration, and the ultimate fate of the system, are yet to be understood. Future multi-wavelength observations, astrometric monitoring (e.g., with \textit{Gaia} and future missions), and theoretical modeling of supernova kicks and binary evolution will be essential to fully unravel the origins and implications of this remarkable cosmic benchmark.

\section*{Acknowledgments}
We would like to thank the anonymous reviewer for his/her professional and detailed suggestions for improving the work and Jue-Ran Niu for providing us with an efficient working environment. H.F.X. acknowledges support from the National Natural Science Foundation of China (NSFC; No. 12303036) and the Scientific and Technological Innovation Programs of Higher Education Institutions in Shanxi (STIP; 2025Q032). All the authors acknowledge the TESS Science team and everyone who has contributed to making the TESS mission possible. We also acknowledge with thanks the variable star observations from the AAVSO International Database contributed by observers worldwide and used in this research.
All the {\it TESS} data used in this paper can be found in \citet{TESS_lc}.

\software{{\tt astropy} \citep{astropy}, {\tt Lightkurve} \citep{lightkurve}, {\tt NumPy} \citep{numpy}, {\tt SciPy} \citep{scipy}, {\tt matplotlib} \citep{matplotlib}, {\tt emcee} \citep{Foreman2013}, {\tt MESA} \citep{Paxton2011, Paxton2013, Paxton2015, Paxton2018, Paxton2019, Jermyn2023}, {\tt GYRE} \citep{Townsend2013}}


\clearpage 
\appendix

\section{Photometric Data and Frequency Analysis}
\label{app:photometry}

\begin{figure}[!htbp]
    \centering
    \includegraphics[width=0.7\textwidth]{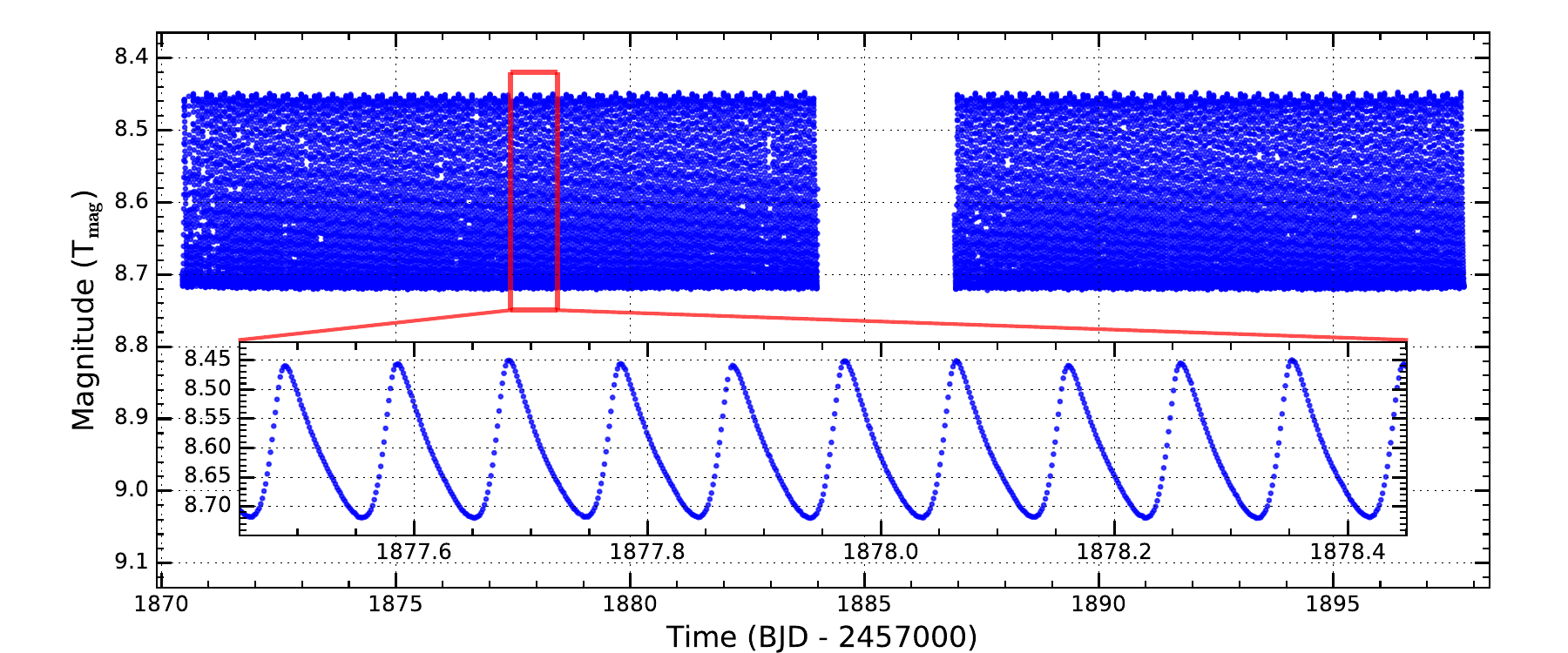}
    \caption{Light curve of BE Lyn from TESS Sector~21. The lower panel shows a one-day zoom for clarity.}
    \label{fig:lc}
\end{figure}

BE~Lyn was observed by the Transiting Exoplanet Survey Satellite (\textit{TESS}) in Sector~21 (January 2020) with a 120-second cadence. We retrieved the Presearch Data Conditioning Simple Aperture Photometry (PDCSAP) flux from the Mikulski Archive for Space Telescopes (MAST) and converted it to magnitudes using the \textit{TESS} magnitude system \citep{smith2012, stumpe2012, stumpe2014, Stassun2019a} (see in Figure~\ref{fig:lc}).

\begin{figure}[!htbp]
    \centering
    \includegraphics[width=0.7\textwidth]{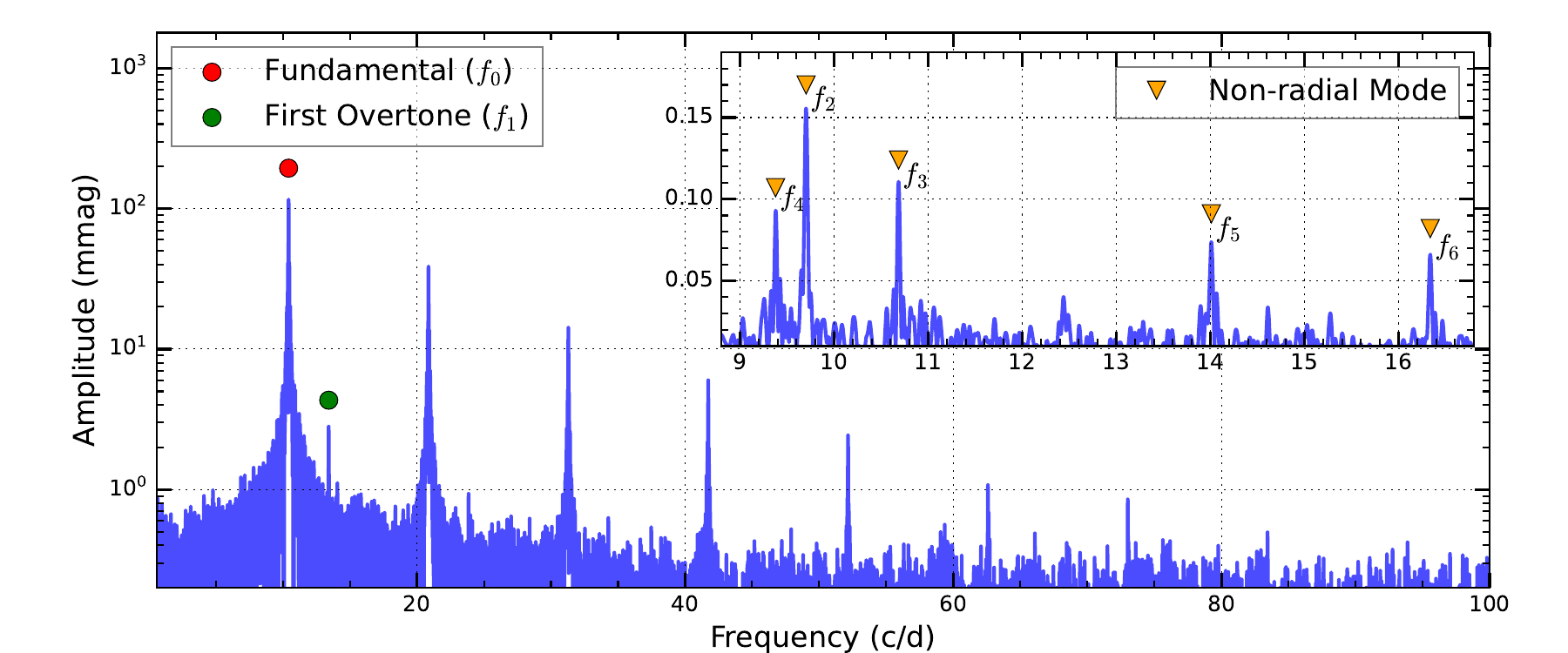}
    \caption{Frequency spectrum of BE Lyn. The main panel ($0$--$100\ \cd$) marks the fundamental ($f_0$, red) and first-overtone ($f_1$, green) modes. The inset highlights the low-amplitude, non-radial mode region ($\sim 9-17\ \cd$) with a linear y-axis.}
    \label{fig:spectra}
\end{figure}

We extracted pulsation frequencies using an iterative prewhitening procedure \citep{Niu2022, Niu2023, Niu2024}. The light curve was modeled as a superposition of sinusoids:
\begin{equation}
    m(t) = m_0 + \sum_i a_i \sin[2\pi(f_i t + \phi_i)],
\end{equation}
where $m_0$ is the mean magnitude, and $a_i$, $f_i$, and $\phi_i$ are the amplitude, frequency, and phase of each component. Frequencies were accepted as significant if their signal-to-noise ratio (S/N) exceeded 5.6 \citep{Zong2016_sdb,Zong2018}. After removing aliases, we identified 32 significant peaks, consisting of 7 independent modes and 25 harmonics or linear combinations (see in Figure~\ref{fig:spectra} and Table~\ref{tab:freqs}).

\begin{deluxetable}{c|ccccc|c}[!htbp]
    \label{tab:freqs}
    \tablecaption{Significant Frequencies Detected in BE Lyn from TESS Sector~21.}
    \tabletypesize{\footnotesize}
    \tablehead{
        \colhead{NO.} & \colhead{$f \ (\cd)$} & \colhead{$\sigma_f \ (\cd)$} & \colhead{$a \ (\mathrm{mmag})$} & \colhead{$\sigma_a \ (\mathrm{mmag})$} & \colhead{S/N} & \colhead{Marks}
    }
    \startdata
				F1  & 10.4309 & 0.0005 & 116 & 3  & 42.4 & $f_0$                       \\
				F2  & 20.8618 & 0.0005 & 38.9  & 0.9  & 42.1 & $2f_0$                        \\
				F3  & 31.2926 & 0.0005 & 14.3  & 0.3  & 42.4 & $3f_0$                       \\
				F4  & 41.7236 & 0.0005 & 6.0   & 0.1  & 44.2 & $4f_0$                        \\
				F5  & 13.4226 & 0.0005 & 2.59   & 0.07  & 37.8 & $f_1$                       \\
				F6  & 52.1544 & 0.0005 & 2.57   & 0.06  & 40.3 & $5f_0$                       \\
				F7  & 62.5852 & 0.0005 & 1.41   & 0.03  & 41.3 & $6f_0$                       \\
				F8  & 23.8534 & 0.0005 & 1.04   & 0.03  & 38.4 & $f_0+f_1$                       \\
				F9  & 73.0156 & 0.0006 & 0.79   & 0.02  & 34.6 & $7f_0$                       \\
				F10 & 2.9914  & 0.0007 & 0.71   & 0.03  & 27.4 & $-f_0+f_1$                       \\
				F11 & 34.2846 & 0.0008 & 0.50   & 0.02  & 27.0 & $2f_0+f_1$                        \\
				F12 & 83.4470 & 0.0006 & 0.47   & 0.01  & 31.8 & $8f_0$                       \\
				F13 & 7.440  & 0.001 & 0.37   & 0.02  & 21.3 &$2f_0-f_1$                        \\
				F14 & 44.7150  & 0.0008 & 0.29   & 0.01  & 24.4& $3f_0+f_1$                        \\
				F15 & 93.8765 & 0.0009 & 0.26   & 0.01  & 23.8 &$9f_0$                       \\
				F16 & 55.147  & 0.001   & 0.16   & 0.01  & 17.7 &  $4f_0+f_1$                      \\
				F17 & 9.701   & 0.003  & 0.15   & 0.02 & 7.2 &   $f_2$                     \\
				F18 & 104.309  & 0.002 & 0.14   & 0.01 & 14.0 & $10f_0$                       \\
				F19 & 17.869  & 0.002  & 0.11   & 0.01 & 10.5 &   $3f_0-f_1$                     \\
				F20 & 65.579  & 0.002  & 0.10   & 0.01  & 10.4  & $5f_0+f_1$                       \\
				F21 & 10.684  & 0.003  & 0.10   & 0.02 & 6.0 & $f_3$ \\ 
				F22 & 9.377  & 0.003   & 0.09    & 0.01 & 6.0 & $f_4$                        \\
				F23 & 114.740  & 0.002 & 0.076  & 0.008 & 9.8 &$11f_0$  \\
				F24 & 14.009  & 0.003  & 0.08   & 0.01  & 6.4 &$f_5$                       \\
				F25 & 19.810  & 0.003  & 0.08  & 0.01 & 6.2  & $f_0+f_4$                       \\
				F26 & 20.133  & 0.003  & 0.07   & 0.01  & 5.9 &   $f_0+f_2$                     \\
				F27 & 21.112  & 0.004  & 0.07   & 0.01  &5.7 &  $f_0+2f_3$                      \\
				F28 & 76.006  & 0.003  & 0.067   & 0.009  &7.5 &   $6f_0+f_1$                     \\
				F29 & 16.336  & 0.003  & 0.07   & 0.01  & 6.8 &  $f_6$                      \\
				F30 & 28.299  & 0.003  & 0.058  & 0.009 & 6.1  & $4f_0-f_1$                       \\
				F31 & 38.729  & 0.003  & 0.052  & 0.008 & 6.1 & $5f_0-f_1$                       \\
				F32 & 86.438  & 0.004  & 0.045   & 0.008  &5.8 &   $7f_0+f_1$                      \\
    \enddata
\end{deluxetable}

\section{Compilation and Processing of Times of Maximum Light}
\label{app:tml_compilation}

The light curve is dominated by the fundamental mode ($f_0$), with the first overtone ($f_1$) contributing less than 3\% of its amplitude \citep{Xue2023}. This allows the classical $O-C$ method \citep{Sterken_2005} to be applied without significant contamination from multi-periodic effects.

We compiled times of maximum light (TML) from three sources: (1) historical literature, (2) the American Association of Variable Star Observers (AAVSO) database, and (3) \textit{TESS} Sector~21 photometry. For AAVSO and \textit{TESS} data, we determined TML by fitting a cubic polynomial to each light-curve maximum, with uncertainties estimated via Monte Carlo simulations.

To ensure temporal consistency, all times were converted to Barycentric Julian Date in the TDB system (BJD$_{\mathrm{TDB}}$). Heliocentric Julian Dates (HJD) were transformed using the tool of \citet{Eastman_2010}, while Julian Dates (JD) were converted using the \texttt{Astropy} time package.

Our final dataset includes 442 TML spanning 39 years (1986--2025), after excluding 16 outliers identified in a preliminary $O-C$ analysis. For TML from \citet{Boonyarak2011} published without uncertainties, we adopted a conservative error of 0.0001 d, consistent with their reported precision and twice the mean uncertainty of our \textit{TESS}-based measurements.

\section{Orbital Modeling via MCMC}
\label{app:orbital_fitting}

The $O-C$ residuals were modeled with a function combining a quadratic term (intrinsic period variation) and a light-travel time effect (LTTE) term for orbital motion \citep{Paparo1988,Xue2020}:
\begin{equation}
  \label{eq:oc-fitted}
  C = \ \mathrm{BJD}_0 + P_0 E + \frac{1}{2} \beta E^2 + A\left[\sqrt{1-e^2}\sin{\phi} \cos{\omega}+\cos{\phi}\sin{\omega}\right],
\end{equation}
where $\phi$ is the eccentric anomaly from Kepler's equation, $\mathrm{BJD}_0$ is the reference epoch, $P_0$ the pulsation period, $\beta$ the linear period variation rate, $A \equiv a\sin i / c$ the projected semi-major axis, $e$ the orbital eccentricity, $\omega$ the argument of periastron, $P_{\mathrm{orb}}$ the orbital period, and $T_0$ the time of periastron passage.

We employed a Markov Chain Monte Carlo (MCMC) approach using the \texttt{emcee} package \citep{Foreman2013} to sample the posterior distributions of all parameters. After chain convergence, we adopted the median of each posterior as the best-fit value, with uncertainties derived from the 5th and 95th percentiles. The best-fit model ($\chi^{2}/\mathrm{d.o.f.} = 35.23$) is shown in Figure~\ref{fig:OC_BELyn}, with parameters listed in Table~\ref{tab:pul_orb_para}.

\section{Asteroseismic Modeling of the Primary Star}
\label{app:asteroseismology}

We estimated the primary star's properties by matching the observed pulsation characteristics to stellar models. Using MESA (Modules for Experiments in Stellar Astrophysics) \citep{Paxton2011, Paxton2013, Paxton2015, Paxton2018, Paxton2019, Jermyn2023}, we computed a grid of evolutionary tracks from the pre-main sequence to the red giant branch, with masses from 1.3 to 2.3 $M_{\odot}$ in steps of 0.01 $M_{\odot}$ and metallicity $Z=0.0034$ (from APOGEE-2 \citep{Abdurrouf2022}, corresponding to [Fe/H] = $-$0.598). At each time step, we calculated adiabatic oscillation frequencies using GYRE \citep{Townsend2013, Townsend2018, Goldstein2020, Sun2023}.
More details about the theoretical calculation can be referred to in \citet{Niu2022} and \citet{Niu2025}.

The optimal model was selected by minimizing $\chi^{2}$ across two observables: the fundamental frequency ($f_0$) and first-overtone frequency ($f_1$). This methodology follows established asteroseismic analyses of HADS \citep{Niu2017, Xue2018, Xue2022}. The best-fit model parameters are given in Table~\ref{tab:primary_para}.

\begin{deluxetable}{c|cc}[!htbp]
    \label{tab:primary_para}
    \tablecaption{Parameters of the best-fit seismic model for BE~Lyn's primary star.}
    \tabletypesize{\footnotesize}
    \tablehead{
        \colhead{Parameter} & \colhead{Best-fit Model} & \colhead{Observed}
    }
    \startdata
    $f_0$ (c/d) & 10.4310 & $10.4309 \pm 0.0005$ \\
    $f_1$ (c/d) & 13.4226 & $13.4226 \pm 0.0005$ \\
    $(1/P_0)(dP_0/dt)$ (yr$^{-1}$) & $8.36 \times 10^{-8}$ & $(1.02^{+0.08}_{-0.08}) \times 10^{-8}$ \\ \hline
    $\log T_{\mathrm{eff}}$ & 3.915 & -- \\
    $\log (L/L_{\odot})$ & 1.391 & -- \\
    $\log g$ & 3.870 & -- \\
    $\log (R/R_{\odot})$ & 0.387 & -- \\
    Age (Gyr) & 1.44 & -- \\
    Mass ($M_{\odot}$) & 1.61 & -- \\
    \enddata
\end{deluxetable}

\section{Dynamical Constraints on the Companion}
\label{app:dynamics}

The companion's nature was inferred from the mass function,
\begin{equation}
\label{eq:mass_fun}
    f(M_2) = \frac{(M_2 \sin i)^3}{(M_1 + M_2)^2} = \frac{4\pi^2}{G P_{\mathrm{orb}}^2} (a \sin i)^3,
\end{equation}
combined with the constraints that the periastron separation $d_{\mathrm{peri}} = a(1-e)(M_1+M_2)/M_2$ must exceed the primary's radius $R_1$. This yields an upper limit on the inclination ($i \lesssim 10.1^\circ$) and a corresponding lower limit on the companion mass ($M_2 \gtrsim 2.5 M_{\odot}$), as shown in Figure~\ref{fig:m2_mass}.

\clearpage
\section{Additional Tables} 
All the TML collected and used in this work are listed in Table \ref{tab:TML}.	

\startlongtable
\begin{deluxetable*}{llcc|llcc|llcc}
\label{tab:TML}
	\tablecaption{Times of maximum light of BE Lyn. The numbers in the fourth, eighth, and twelfth columns indicate the different data sources: (1) \citet{Oja_1987}; (2) \citet{Rodriguez_1990}; (3) \citet{Liu_1991}; (4) \citet{Wunder_1992}; (5) \citet{Tang_1992}; (6) \citet{Liu_1994}; (7) \citet{Kiss_1995}; (8) \citet{Derekas_2003}; (9) \citet{Fu_2005}; (10) \citet{Szakats_2008}; (11) \citet{Hubscher_2005};(12) \citet{Hubscher_2005a}; (13) \citet{Hubscher_2006}; (14) \citet{Hubscher_2005b}; (15) \citet{Klingenberg_2006}; (16) \citet{Boonyarak2011}; (17) \citet{Hubscher_2007}; (18) \citet{Hubscher_2009}; (19) AAVSO; (20) \citet{Hubscher_2013}; (21) \citet{Pena2015}; (22) \citet{Hubscher_2014}; (23) \citet{Hubscher_2017}; (24) TESS; (25) \citet{Pagel_2022}; (26) \citet{Pagel_2021}; (27) \citet{Pagel_2023}.}
	\tablehead{
		\colhead{TML($\BJD$)} & \colhead{Error} & \colhead{Detector} & \colhead{Source} & \colhead{TML($\BJD$)} & \colhead{Error} & \colhead{Detector} & \colhead{Source} &\colhead{TML($\BJD$)} & \colhead{Error} & \colhead{Detector} & \colhead{Source} 
	} 
	\startdata
	2446498.3387 & 0.0002 & pe & 1 & 2454433.6515 & 0.0008 & CCD & 18 & 2458882.47489 & 0.00004 & CCD & 24 \\
	2446507.3500 & 0.0002 & pe & 1 & 2454837.356\tablenotemark{a} & 0.001 & CCD & 18 & 2458882.57111 & 0.00004 & CCD & 24 \\
	2446507.4466 & 0.0002 & pe & 1 & 2454858.6447\tablenotemark{*} & 0.0001 & CCD & 16 & 2458882.66680 & 0.00003 & CCD & 24 \\
	2446508.4052 & 0.0002 & pe & 1 & 2454871.3891 & 0.0001 & CCD & 16 & 2458882.76233 & 0.00004 & CCD & 24 \\
	2446509.4600 & 0.0002 & pe & 1 & 2454946.3599 & 0.0001 & CCD & 16 & 2458882.85851 & 0.00004 & CCD & 24 \\
	2446510.4184 & 0.0002 & pe & 1 & 2455309.6105 & 0.0002 & CCD & 19 & 2458882.95462 & 0.00004 & CCD & 24 \\
	2446524.4151 & 0.0002 & pe & 1 & 2455646.3031 & 0.0006 & CCD & 20 & 2458883.04995 & 0.00004 & CCD & 24 \\
	2446950.4592 & 0.0001 & pe & 2 & 2455646.3992 & 0.0005 & CCD & 20 & 2458883.14582 & 0.00004 & CCD & 24 \\
	2446951.4185 & 0.0001 & pe & 2 & 2455968.3297 & 0.0007 & CCD & 20 & 2458883.24205 & 0.00004 & CCD & 24 \\
	2447115.6426 & 0.0001 & pe & 2 & 2456002.363 & 0.002 & CCD & 20 & 2458883.33778 & 0.00003 & CCD & 24 \\
	2447118.6144 & 0.0001 & pe & 2 & 2456046.465 & 0.002 & CCD & 20 & 2458883.43331 & 0.00003 & CCD & 24 \\
	2447118.7104 & 0.0002 & pe & 2 & 2456050.393 & 0.004 & CCD & 20 & 2458883.52961 & 0.00004 & CCD & 24 \\
	2447121.6822 & 0.0001 & pe & 2 & 2456347.780 & 0.002 & CCD & 21 & 2458883.62569 & 0.00003 & CCD & 24 \\
	2447219.5648 & 0.0002 & pe & 2 & 2456347.876\tablenotemark{a} & 0.002 & CCD & 21 & 2458883.72091 & 0.00003 & CCD & 24 \\
	2447265.1032 & 0.0001 & pe & 3 & 2456360.820 & 0.002 & CCD & 21 & 2458883.81699 & 0.00004 & CCD & 24 \\
	2447486.3699 & 0.0001 & pe & 3 & 2456371.366 & 0.003 & DSLR & 22 & 2458883.91322 & 0.00004 & CCD & 24 \\
	2447488.2859 & 0.0001 & pe & 3 & 2456374.815 & 0.002 & CCD & 21 & 2458886.98108 & 0.00003 & CCD & 24 \\
	2447488.3818 & 0.0001 & pe & 3 & 2456376.638 & 0.002 & CCD & 21 & 2458887.07648 & 0.00003 & CCD & 24 \\
	2447489.2457 & 0.0001 & pe & 3 & 2456376.733 & 0.002 & CCD & 21 & 2458887.17252 & 0.00004 & CCD & 24 \\
	2447489.3419 & 0.0001 & pe & 3 & 2456657.440\tablenotemark{b} & 0.001 & CCD & 23 & 2458887.26876 & 0.00004 & CCD & 24 \\
	2447493.1762 & 0.0001 & pe & 3 & 2457454.955\tablenotemark{*} & 0.002 & DSLR & 19 & 2458887.36421 & 0.00003 & CCD & 24 \\
	2447493.2710 & 0.0001 & pe & 3 & 2457458.3346 & 0.0007 & CCD & 23 & 2458887.45982 & 0.00003 & CCD & 24 \\
	2447542.2612 & 0.0001 & pe & 3 & 2457458.4306 & 0.0006 & CCD & 23 & 2458887.55606 & 0.00003 & CCD & 24 \\
	2447544.2737 & 0.0001 & pe & 3 & 2458847.3857\tablenotemark{c} & 0.0005 & DSLR & 19 & 2458887.65191 & 0.00003 & CCD & 24 \\
	2447544.3702 & 0.0001 & pe & 3 & 2458847.4825\tablenotemark{c} & 0.0003 & DSLR & 19 & 2458887.74747 & 0.00003 & CCD & 24 \\
	2447551.4654 & 0.0001 & pe & 2 & 2458847.5786\tablenotemark{c} & 0.0003 & DSLR & 19 & 2458887.84355 & 0.00004 & CCD & 24 \\
	2447551.5613 & 0.0001 & pe & 2 & 2458847.6726\tablenotemark{c} & 0.0005 & DSLR & 19 & 2458887.93986 & 0.00004 & CCD & 24 \\
	2447551.6567 & 0.0001 & pe & 2 & 2458847.7696\tablenotemark{c} & 0.0002 & DSLR & 19 & 2458888.03535 & 0.00004 & CCD & 24 \\
	2447553.5743 & 0.0001 & pe & 2 & 2458870.49132 & 0.00004 & CCD & 24 & 2458888.13094 & 0.00004 & CCD & 24 \\
	2447553.6702 & 0.0001 & pe & 2 & 2458870.58827 & 0.00003 & CCD & 24 & 2458888.22723 & 0.00004 & CCD & 24 \\
	2447627.0095 & 0.0001 & pe & 3 & 2458870.68287 & 0.00004 & CCD & 24 & 2458888.32305 & 0.00003 & CCD & 24 \\
	2447634.0087 & 0.0001 & pe & 3 & 2458870.77880 & 0.00004 & CCD & 24 & 2458888.41851 & 0.00004 & CCD & 24 \\
	2448347.3735 & 0.0001 & pe & 4 & 2458870.87510 & 0.00003 & CCD & 24 & 2458888.51457 & 0.00004 & CCD & 24 \\
	2448347.4686 & 0.0001 & pe & 4 & 2458870.97065 & 0.00004 & CCD & 24 & 2458888.61086 & 0.00004 & CCD & 24 \\
	2448357.3451 & 0.0001 & pe & 4 & 2458871.06612 & 0.00004 & CCD & 24 & 2458888.70635 & 0.00003 & CCD & 24 \\
	2448358.3995 & 0.0001 & pe & 4 & 2458871.16239 & 0.00004 & CCD & 24 & 2458888.80202 & 0.00003 & CCD & 24 \\
	2448359.3564 & 0.0001 & pe & 4 & 2458871.25848 & 0.00004 & CCD & 24 & 2458888.89837 & 0.00004 & CCD & 24 \\
	2448367.5067 & 0.0001 & pe & 4 & 2458871.35387 & 0.00003 & CCD & 24 & 2458888.99415 & 0.00004 & CCD & 24 \\
	2448624.3407 & 0.0001 & pe & 4 & 2458871.44980 & 0.00004 & CCD & 24 & 2458889.08950 & 0.00004 & CCD & 24 \\
	2448707.0770 & 0.0001 & pe & 5 & 2458871.54610 & 0.00004 & CCD & 24 & 2458889.18563 & 0.00004 & CCD & 24 \\
	2448710.0480 & 0.0001 & pe & 5 & 2458871.64164 & 0.00004 & CCD & 24 & 2458889.28186 & 0.00004 & CCD & 24 \\
	2448710.1441 & 0.0001 & pe & 5 & 2458871.73743 & 0.00004 & CCD & 24 & 2458889.37750 & 0.00004 & CCD & 24 \\
	2448710.2397 & 0.0001 & pe & 5 & 2458871.83361 & 0.00004 & CCD & 24 & 2458889.47307 & 0.00004 & CCD & 24 \\
	2448715.1294 & 0.0001 & pe & 5 & 2458871.92941 & 0.00003 & CCD & 24 & 2458889.56947 & 0.00004 & CCD & 24 \\
	2448715.2260 & 0.0001 & pe & 5 & 2458872.02491 & 0.00003 & CCD & 24 & 2458889.66535 & 0.00004 & CCD & 24 \\
	2449018.0775 & 0.0001 & pe & 6 & 2458872.12093 & 0.00004 & CCD & 24 & 2458889.76080 & 0.00004 & CCD & 24 \\
	2449018.1733 & 0.0001 & pe & 6 & 2458872.21706 & 0.00004 & CCD & 24 & 2458889.85684 & 0.00003 & CCD & 24 \\
	2449018.2692 & 0.0001 & pe & 6 & 2458872.31280 & 0.00004 & CCD & 24 & 2458889.95305 & 0.00004 & CCD & 24 \\
	2449018.3651 & 0.0001 & pe & 6 & 2458872.40836 & 0.00003 & CCD & 24 & 2458890.04850 & 0.00003 & CCD & 24 \\
	2449019.0359 & 0.0001 & pe & 6 & 2458872.50464 & 0.00004 & CCD & 24 & 2458890.14419 & 0.00004 & CCD & 24 \\
	2449021.0507 & 0.0001 & pe & 6 & 2458872.60063 & 0.00003 & CCD & 24 & 2458890.24059 & 0.00004 & CCD & 24 \\
	2449022.0086 & 0.0001 & pe & 6 & 2458872.69606 & 0.00004 & CCD & 24 & 2458890.33633 & 0.00004 & CCD & 24 \\
	2449030.0612 & 0.0001 & pe & 6 & 2458872.79194 & 0.00004 & CCD & 24 & 2458890.43185 & 0.00004 & CCD & 24 \\
	2449031.0199 & 0.0001 & pe & 6 & 2458872.88832 & 0.00004 & CCD & 24 & 2458890.52798 & 0.00003 & CCD & 24 \\
	2449400.0222 & 0.0001 & pe & 6 & 2458872.98382 & 0.00003 & CCD & 24 & 2458890.62419 & 0.00004 & CCD & 24 \\
	2449400.1180 & 0.0001 & pe & 6 & 2458873.07934 & 0.00004 & CCD & 24 & 2458890.71956 & 0.00003 & CCD & 24 \\
	2449401.0760 & 0.0001 & pe & 6 & 2458873.17571 & 0.00004 & CCD & 24 & 2458890.81532 & 0.00003 & CCD & 24 \\
	2449401.1720 & 0.0001 & pe & 6 & 2458873.27175 & 0.00004 & CCD & 24 & 2458890.91166 & 0.00004 & CCD & 24 \\
	2449413.0600 & 0.0001 & pe & 6 & 2458873.36714 & 0.00004 & CCD & 24 & 2458891.00736 & 0.00003 & CCD & 24 \\
	2449413.1560 & 0.0001 & pe & 6 & 2458873.46306 & 0.00004 & CCD & 24 & 2458891.10285 & 0.00003 & CCD & 24 \\
	2449416.1279 & 0.0001 & pe & 6 & 2458873.55933 & 0.00004 & CCD & 24 & 2458891.19901 & 0.00004 & CCD & 24 \\
	2449749.4657 & 0.0002 & pe & 7 & 2458873.65484 & 0.00003 & CCD & 24 & 2458891.29526 & 0.00004 & CCD & 24 \\
	2449749.5616 & 0.0002 & pe & 7 & 2458873.75062 & 0.00003 & CCD & 24 & 2458891.39060 & 0.00003 & CCD & 24 \\
	2449749.6575 & 0.0002 & pe & 7 & 2458873.84678 & 0.00004 & CCD & 24 & 2458891.48640 & 0.00004 & CCD & 24 \\
	2449754.3549 & 0.0004 & pe & 7 & 2458873.94276 & 0.00003 & CCD & 24 & 2458891.58276 & 0.00004 & CCD & 24 \\
	2449762.4077 & 0.0004 & pe & 8 & 2458874.03815 & 0.00004 & CCD & 24 & 2458891.67854 & 0.00004 & CCD & 24 \\
	2450114.5373 & 0.0005 & pe & 8 & 2458874.13411 & 0.00004 & CCD & 24 & 2458891.77394 & 0.00004 & CCD & 24 \\
	2450139.4632 & 0.0004 & pe & 8 & 2458874.23035 & 0.00004 & CCD & 24 & 2458891.87012 & 0.00004 & CCD & 24 \\
	2450140.4216 & 0.0002 & pe & 8 & 2458874.32609 & 0.00003 & CCD & 24 & 2458891.96628 & 0.00004 & CCD & 24 \\
	2450500.5078 & 0.0002 & pe & 8 & 2458874.42166 & 0.00003 & CCD & 24 & 2458892.06157 & 0.00003 & CCD & 24 \\
	2450844.3916 & 0.0001 & pe & 8 & 2458874.51795 & 0.00004 & CCD & 24 & 2458892.15742 & 0.00003 & CCD & 24 \\
	2450844.4874 & 0.0001 & pe & 8 & 2458874.61393 & 0.00004 & CCD & 24 & 2458892.25382 & 0.00004 & CCD & 24 \\
	2451205.3404 & 0.0002 & pe & 9 & 2458874.70925 & 0.00004 & CCD & 24 & 2458892.34951 & 0.00004 & CCD & 24 \\
	2451205.5322 & 0.0001 & pe & 8 & 2458874.80541 & 0.00004 & CCD & 24 & 2458892.44507 & 0.00004 & CCD & 24 \\
	2451205.6280 & 0.0001 & pe & 8 & 2458874.90155 & 0.00004 & CCD & 24 & 2458892.54120 & 0.00003 & CCD & 24 \\
	2451205.7239 & 0.0001 & pe & 8 & 2458874.99696 & 0.00004 & CCD & 24 & 2458892.63732 & 0.00003 & CCD & 24 \\
	2451207.3549 & 0.0001 & pe & 9 & 2458875.09267 & 0.00003 & CCD & 24 & 2458892.73273 & 0.00004 & CCD & 24 \\
	2451209.2717 & 0.0001 & pe & 9 & 2458875.18902 & 0.00004 & CCD & 24 & 2458892.82861 & 0.00004 & CCD & 24 \\
	2451209.3669 & 0.0001 & pe & 9 & 2458875.28492 & 0.00003 & CCD & 24 & 2458892.92495 & 0.00004 & CCD & 24 \\
	2451237.2645 & 0.0002 & pe & 8 & 2458875.38040 & 0.00004 & CCD & 24 & 2458893.02060 & 0.00004 & CCD & 24 \\
	2451237.3603 & 0.0002 & pe & 8 & 2458875.47644 & 0.00003 & CCD & 24 & 2458893.11598 & 0.00004 & CCD & 24 \\
	2451237.4562 & 0.0002 & pe & 8 & 2458875.57261 & 0.00003 & CCD & 24 & 2458893.21221 & 0.00005 & CCD & 24 \\
	2451237.5521 & 0.0002 & pe & 8 & 2458875.66818 & 0.00003 & CCD & 24 & 2458893.30843 & 0.00004 & CCD & 24 \\
	2451239.3737 & 0.0003 & pe & 8 & 2458875.76375 & 0.00004 & CCD & 24 & 2458893.40388 & 0.00003 & CCD & 24 \\
	2451239.4696 & 0.0003 & pe & 8 & 2458875.85999 & 0.00004 & CCD & 24 & 2458893.49967 & 0.00004 & CCD & 24 \\
	2451240.4287 & 0.0001 & pe & 8 & 2458875.95596 & 0.00004 & CCD & 24 & 2458893.59602 & 0.00004 & CCD & 24 \\
	2451240.5246 & 0.0001 & pe & 8 & 2458876.05147 & 0.00003 & CCD & 24 & 2458893.69190 & 0.00004 & CCD & 24 \\
	2451240.6205 & 0.0001 & pe & 8 & 2458876.14746 & 0.00004 & CCD & 24 & 2458893.78726 & 0.00004 & CCD & 24 \\
	2451243.3041 & 0.0002 & pe & 8 & 2458876.24366 & 0.00004 & CCD & 24 & 2458893.88343 & 0.00004 & CCD & 24 \\
	2451243.4000 & 0.0002 & pe & 8 & 2458876.33924 & 0.00003 & CCD & 24 & 2458893.97944 & 0.00004 & CCD & 24 \\
	2451246.2764 & 0.0001 & pe & 8 & 2458876.43493 & 0.00004 & CCD & 24 & 2458894.07487 & 0.00004 & CCD & 24 \\
	2451246.3723 & 0.0001 & pe & 8 & 2458876.53122 & 0.00004 & CCD & 24 & 2458894.17079 & 0.00004 & CCD & 24 \\
	2451246.4681 & 0.0001 & pe & 8 & 2458876.62707 & 0.00003 & CCD & 24 & 2458894.26713 & 0.00004 & CCD & 24 \\
	2451293.3486 & 0.0001 & pe & 8 & 2458876.72311 & 0.00006 & CCD & 24 & 2458894.36280 & 0.00004 & CCD & 24 \\
	2451293.4444 & 0.0001 & pe & 8 & 2458876.81860 & 0.00004 & CCD & 24 & 2458894.45822 & 0.00004 & CCD & 24 \\
	2451311.4677 & 0.0002 & pe & 8 & 2458876.91469 & 0.00004 & CCD & 24 & 2458894.55453 & 0.00004 & CCD & 24 \\
	2451313.4810 & 0.0002 & pe & 8 & 2458877.01031 & 0.00003 & CCD & 24 & 2458894.65061 & 0.00004 & CCD & 24 \\
	2451585.3671 & 0.0002 & pe & 8 & 2458877.10601 & 0.00004 & CCD & 24 & 2458894.74601 & 0.00003 & CCD & 24 \\
	2451929.4429 & 0.0002 & pe & 8 & 2458877.20231 & 0.00004 & CCD & 24 & 2458894.84178 & 0.00003 & CCD & 24 \\
	2451929.5387 & 0.0002 & pe & 8 & 2458877.29814 & 0.00004 & CCD & 24 & 2458894.93818 & 0.00004 & CCD & 24 \\
	2451965.2983 & 0.0001 & CCD & 8 & 2458877.39364 & 0.00003 & CCD & 24 & 2458895.03386 & 0.00004 & CCD & 24 \\
	2451965.3941 & 0.0001 & CCD & 8 & 2458877.48958 & 0.00004 & CCD & 24 & 2458895.12929 & 0.00004 & CCD & 24 \\
	2451965.4903 & 0.0001 & CCD & 8 & 2458877.58585 & 0.00003 & CCD & 24 & 2458895.22565 & 0.00004 & CCD & 24 \\
	2451985.3347 & 0.0002 & CCD & 8 & 2458877.68135 & 0.00003 & CCD & 24 & 2458895.32171 & 0.00003 & CCD & 24 \\
	2451985.5264 & 0.0002 & CCD & 8 & 2458877.77708 & 0.00004 & CCD & 24 & 2458895.41708 & 0.00004 & CCD & 24 \\
	2452252.5198\tablenotemark{*} & 0.0001 & CCD & 8 & 2458877.87348 & 0.00004 & CCD & 24 & 2458895.51297 & 0.00004 & CCD & 24 \\
	2452371.4010 & 0.0002 & CCD & 8 & 2458877.96917 & 0.00004 & CCD & 24 & 2458895.60936 & 0.00004 & CCD & 24 \\
	2452371.4969 & 0.0002 & CCD & 8 & 2458878.06464 & 0.00004 & CCD & 24 & 2458895.70493 & 0.00004 & CCD & 24 \\
	2452396.3273 & 0.0001 & CCD & 8 & 2458878.16078 & 0.00004 & CCD & 24 & 2458895.80042 & 0.00004 & CCD & 24 \\
	2452396.4232 & 0.0001 & CCD & 8 & 2458878.25694 & 0.00004 & CCD & 24 & 2458895.89673 & 0.00004 & CCD & 24 \\
	2452981.5147 & 0.0001 & CCD & 10 & 2458878.35252 & 0.00004 & CCD & 24 & 2458895.99269 & 0.00004 & CCD & 24 \\
	2452981.6106 & 0.0001 & CCD & 10 & 2458878.44826 & 0.00004 & CCD & 24 & 2458896.08806 & 0.00004 & CCD & 24 \\
	2453056.485\tablenotemark{a} & 0.001 & pe & 11 & 2458878.54457 & 0.00004 & CCD & 24 & 2458896.18406 & 0.00004 & CCD & 24 \\
	2453056.582\tablenotemark{a} & 0.001 & pe & 11 & 2458878.64034 & 0.00003 & CCD & 24 & 2458896.28041 & 0.00004 & CCD & 24 \\
	2453094.3522 & 0.0008 & pe & 12 & 2458878.73583 & 0.00003 & CCD & 24 & 2458896.37596 & 0.00004 & CCD & 24 \\
	2453094.371\tablenotemark{b*} & 0.007 & pe & 11 & 2458878.83192 & 0.00003 & CCD & 24 & 2458896.47141 & 0.00004 & CCD & 24 \\
	2453094.4474 & 0.0008 & pe & 12 & 2458878.92801 & 0.00004 & CCD & 24 & 2458896.56775 & 0.00004 & CCD & 24 \\
	2453111.323\tablenotemark{b} & 0.002 & pe & 11 & 2458879.02343 & 0.00004 & CCD & 24 & 2458896.66375 & 0.00004 & CCD & 24 \\
	2453111.419\tablenotemark{b} & 0.003 & pe & 11 & 2458879.11923 & 0.00004 & CCD & 24 & 2458896.75910 & 0.00004 & CCD & 24 \\
	2453111.517\tablenotemark{b*} & 0.003 & pe & 11 & 2458879.21571 & 0.00004 & CCD & 24 & 2458896.85515 & 0.00004 & CCD & 24 \\
	2453118.4162 & 0.0001 & CCD & 10 & 2458879.31139 & 0.00004 & CCD & 24 & 2458896.95144 & 0.00004 & CCD & 24 \\
	2453349.462 & 0.001 & CCD & 13 & 2458879.40675 & 0.00004 & CCD & 24 & 2458897.04707 & 0.00003 & CCD & 24 \\
	2453349.558 & 0.001 & CCD & 13 & 2458879.50304 & 0.00004 & CCD & 24 & 2458897.14258 & 0.00003 & CCD & 24 \\
	2453349.653 & 0.001 & CCD & 13 & 2458879.59920 & 0.00004 & CCD & 24 & 2458897.23891 & 0.00004 & CCD & 24 \\
	2453352.439\tablenotemark{*} & 0.001 & vis & 14 & 2458879.69459 & 0.00003 & CCD & 24 & 2458897.33481 & 0.00003 & CCD & 24 \\
	2453410.542\tablenotemark{b*} & 0.005 & pe & 12 & 2458879.79036 & 0.00003 & CCD & 24 & 2458897.43031 & 0.00004 & CCD & 24 \\
	2453416.2791\tablenotemark{a*} & 0.0005 & CCD & 15 & 2458879.88672 & 0.00004 & CCD & 24 & 2458897.52626 & 0.00004 & CCD & 24 \\
	2453416.3759\tablenotemark{a} & 0.0005 & CCD & 15 & 2458879.98240 & 0.00003 & CCD & 24 & 2458897.62254 & 0.00004 & CCD & 24 \\
	2453443.6034\tablenotemark{a} & 0.0004 & CCD & 15 & 2458880.07791 & 0.00003 & CCD & 24 & 2458897.71815 & 0.00004 & CCD & 24 \\
	2453472.372\tablenotemark{b*} & 0.005 & pe & 12 & 2458880.17417 & 0.00004 & CCD & 24 & 2458954.378\tablenotemark{c} & 0.003 & CCD & 25 \\
	2453472.465\tablenotemark{b*} & 0.002 & pe & 12 & 2458880.27026 & 0.00004 & CCD & 24 & 2458955.35\tablenotemark{c*} & 0.01 & DSLR & 26 \\
	2453789.1190 & 0.0001 & CCD & 16 & 2458880.36559 & 0.00003 & CCD & 24 & 2458961.37666\tablenotemark{c} & 0.00009 & DSLR & 19 \\
	2453789.2140 & 0.0001 & CCD & 16 & 2458880.46137 & 0.00003 & CCD & 24 & 2458961.462\tablenotemark{*} & 0.001 & DSLR & 26 \\
	2453790.0752 & 0.0001 & CCD & 16 & 2458880.55773 & 0.00004 & CCD & 24 & 2458965.403\tablenotemark{c} & 0.001 & CCD & 25 \\
	2453791.0344 & 0.0001 & CCD & 16 & 2458880.65346 & 0.00004 & CCD & 24 & 2458966.5566\tablenotemark{c*} & 0.0001& DSLR & 19 \\
	2453791.1307 & 0.0001 & CCD & 16 & 2458880.74899 & 0.00003 & CCD & 24 & 2458967.51176\tablenotemark{c} & 0.00009 & DSLR & 19 \\
	2453792.0897 & 0.0001 & CCD & 16 & 2458880.84519 & 0.00004 & CCD & 24 & 2458975.4703\tablenotemark{c} & 0.0003 & DSLR & 19 \\
	2453792.1859 & 0.0001 & CCD & 16 & 2458880.94130 & 0.00003 & CCD & 24 & 2459277.5562\tablenotemark{*} & 0.0002 & CCD & 19 \\
	2453793.0473 & 0.0001 & CCD & 16 & 2458881.03670 & 0.00004 & CCD & 24 & 2459297.399\tablenotemark{c} & 0.004 & DSLR & 25 \\
	2453793.1448 & 0.0001 & CCD & 16 & 2458881.13260 & 0.00004 & CCD & 24 & 2459303.345\tablenotemark{c*} & 0.004 & DSLR & 25 \\
	2453798.3189\tablenotemark{a*} & 0.0002 & CCD & 15 & 2458881.22889 & 0.00004 & CCD & 24 & 2459689.410\tablenotemark{c} & 0.004 & DSLR & 27 \\
	2454027.5469 & 0.0001 & CCD & 10 & 2458881.32458 & 0.00004 & CCD & 24 & 2459690.3667\tablenotemark{c} & 0.0004 & DSLR & 19 \\
	2454027.6428 & 0.0001 & CCD & 10 & 2458881.42006 & 0.00003 & CCD & 24 & 2459990.4390\tablenotemark{c} & 0.0008 & CCD & 19 \\
	2454028.5056 & 0.0001 & CCD & 10 & 2458881.51638 & 0.00004 & CCD & 24 & 2459990.5347\tablenotemark{c} & 0.0009 & CCD & 19 \\
	2454028.6015 & 0.0001 & CCD & 10 & 2458881.61236 & 0.00003 & CCD & 24 & 2460049.3993\tablenotemark{c} & 0.0001 & DSLR & 19 \\
	2454036.4631 & 0.0002 & CCD & 10 & 2458881.70772 & 0.00004 & CCD & 24 & 2460254.8461\tablenotemark{c} & 0.0002 & DSLR & 19 \\
	2454036.5588 & 0.0002 & CCD & 10 & 2458881.80368 & 0.00004 & CCD & 24 & 2460700.640\tablenotemark{c} & 0.001 & CCD & 19 \\
	2454036.6546 & 0.0002 & CCD & 10 & 2458881.89996 & 0.00004 & CCD & 24 & 2460708.5016\tablenotemark{c} & 0.0007 & CCD & 19 \\
	2454039.4346 & 0.0001 & CCD & 10 & 2458881.99566 & 0.00003 & CCD & 24 & 2460714.542\tablenotemark{c} & 0.001 & CCD & 19 \\
	2454039.5305 & 0.0001 & CCD & 10 & 2458882.09110 & 0.00004 & CCD & 24 & 2460783.5680\tablenotemark{c} & 0.0006 & DSLR & 19 \\
	2454039.6263 & 0.0001 & CCD & 10 & 2458882.18743 & 0.00004 & CCD & 24 & 2460783.6660\tablenotemark{c} & 0.0007 & DSLR & 19 \\
	2454221.3950 & 0.0009 & CCD & 17 & 2458882.28342 & 0.00003 & CCD & 24 & 2460805.4251\tablenotemark{c} & 0.0002 & DSLR & 19 \\
	2454222.4498 & 0.0008 & CCD & 17 & 2458882.37876 & 0.00004 & CCD & 24 & & & & \\
\enddata
	\tablecomments{ 
	\tablenotetext{a}{These TML are observed without filter.}
	\tablenotetext{b}{These TML are observed by IR cut-off filter.}
	\tablenotetext{c}{These TML are observed by the TG band (DSLR green-channel).}
	\tablenotetext{*}{These TML are excluded in the global fitting because of the significant deviations in the preliminary $O-C$ analysis.}
	 }
\end{deluxetable*}

\end{CJK*}
\end{document}